\documentclass[12pt]{article}
\usepackage[dvips]{graphicx}

\begin{document}

\title{
Perturbative
theory approaches to the metastable phase decay}
\author{V Kurasov}
\date{ }

\maketitle

\begin{abstract}
The perturbative theory of the nucleation kinetics is
analyzed. A new improvement is suggested and compared with
numerical calculations.
\end{abstract}

\section{Introduction}

The global kinetics of the nucleation process was the subject of
interest during the last few decades. Various approaches to the
description of nucleation kinetics were elaborated in different
external conditions \cite{1} - \cite{8} for different systems.
One of the most popular  is nowadays the perturbation theory approach
presented in \cite{4}. It was formulated for the external
conditions of the "decay" type: at the initial  moment of time the
initial supersaturation is created in the system by some external
action and later no further external influence can be observed -
the system evolution occurs only due to internal processes of the
droplets formation and the vapor consumption by the growing droplets.
This type of external conditions is rather spread both in
the
experimental investigations and in the theoretical descriptions. The
reason is evident: on one hand the amplitude value of
supersaturation is regulated by an external influence, on the other
hand the form of the back side of the droplets size spectrum and
characteristic duration of the nucleation period is governed by
internal process of droplets formation and growth which gives
information about the nucleation origin.

Here we shall analyze
the perturbation theory \cite{4}
in kinetics of the metastable phase decay. The structure of our
analysis will be the following:
\begin{itemize}
\item
At first we shall see that the recipe given by the perturbation
theory leads to the non-uniform decompositions
\item
Then it will be shown that the perturbation approach in the first
approximation (it is rather difficult to speak about the further
approximations because even the first approximation can not be
calculated analytically up to the very end) can be treated as the
monodisperse approximation
\item
A precise  solution  of evolution equations
will be presented and it
will be compared with the perturbation theory approach. It will be
shown that the error of the perturbation theory approach will be
essential
\item
A new version of monodisperse approximation has been proposed.
This approximation is much more accurate than that given by the
perturbation theory
\end{itemize}

\section{Reduction of the balance equation}

The value of supersaturation $\zeta$ is defined as the ratio
$$
\zeta = \frac{n}{n_{\infty}} - 1
$$
where $n$ is the number density of molecules in mother phase and
$n_{\infty}$ is the number density of molecules in the saturated
mother phase. Initial value of supersaturation is marked as
$\zeta_0$.

The balance equation in \cite{4} can be written  as
$$
\frac{\zeta_0}{\zeta(t) } - 1 =  A \int_0^{\infty} \rho^k
g(\rho,t) d\rho
$$
for the distribution
function $g(\rho,t)$ which satisfies the continuity  equation
$$
\frac{ \partial g(\rho,t)} {\partial t } =  -  v(t)
\frac{ \partial g(\rho,t)} {\partial \rho }
$$
with initial condition
$$
g(0,t) = \frac{I(\zeta(t))}{v(t)}
$$
Here
$\rho
$ is the "size" of the embryo which grows with velocity
$$
v =\frac{\zeta}{t_0}
$$
independent on $\rho$, $A$, $k$ and $t_0$ are some parameters.
Then the number of molecules in the embryo of the "size" $\rho$
will be proportional to $\rho^k$ which explains the sense of the
first equation as the balance one.

One can easily reduce this system of equations to
\begin{equation}\label{1}
\frac{\zeta_0}{\zeta(t) } -1 =  A \int_0^{t} (t-t')^k
\frac{I(\zeta(t'))}{v(t')}
 dt'
 \end{equation}
with slightly another value of parameter $A$.

The case $k=0$ is extracted  by the possibility of the analytical
solution of equation (\ref{1}) which can be reduced to the first
order differential equation
\begin{equation}\label{2}
 \zeta_0 \frac{d}{dt} \zeta(t)   =   - A \zeta(t)^2
\frac{I(\zeta(t))}{v(t)}
 \end{equation}
with an evident integration.
So, we shall discuss this case later as well as the case of small
$k \ll 1$ which can be solved on the base of the solution at
$k=0$.

\section{Singular terms in decompositions}

Having extracted the small parameter
$$
\epsilon = (\zeta_0 \frac{d H}{d\zeta})^{-1}
$$
where $H$ is the height of activation barrier in units $k_B T$,
one can easily see that it approximately equals to the inverse number of
molecules in a critical cluster. This parameter will be the small
parameter of the perturbation theory.

Certainly, after the calculation of all terms in decompositions of
the perturbation theory approach and their summation one can get
something accurate. The problem is to get concrete results in
frames of approximations which can be analytically calculated at
least in main features. But already in the first approximation of
the perturbation theory there appeared an auxiliary
function $\varphi_k$ given by equation
$$
\frac{d \varphi_k}{dx} = \exp(-x^k \varphi_k)
$$
$$\varphi_k(0) = 0
$$
which does not allow analytical solution. So, already the first
approximation (and all other ones) can not be calculated
analytically.

From the last remark it follows that
it is necessary  to have in the first approximation
already a good approximation for the real solution. Nevertheless
it will be shown here that the first approximation is not accurate
enough. Although it was announced in \cite{4} that the decomposition  goes on
the small parameter similar to the inverse number of molecules in droplet,
the first approximation is far from the real solution. This occurs due to
the non-uniform character of decompositions. Namely in the balance
equation (12) in \cite{4} one can see that $\rho^k$ transforms
into
\begin{equation}\label{3}
(1-\frac{\epsilon x}{\zeta_0 \tau} + \epsilon w_1 + ...)^k
\end{equation}
with dimensionless variables
$$
x= \frac{\zeta_0 t /t_0 - \rho}{\epsilon}
$$
playing the role of shifted size
and
$$
\tau = \frac{t}{t_0} - \epsilon w_1
$$
playing the role of time. All other parameters can be found in
\cite{4}. It is evident that the r.h.s. is irregular when $\tau$
goes to zero. From the first point of view the limit
$\tau \rightarrow 0 $
corresponds to the negligible part of formation of droplets.
But the careful analysis (see the next section) shows that the nucleation
period duration has the relative smallness less than $\epsilon$ being
compared with the imaginary time of consumption of the main part of surplus
substance.

\section{Relative smallness of the nucleation period duration}

For all  $k$ we see that
\begin{equation}
-\frac{\zeta_0}{\zeta^2(t) }\frac{d \zeta}{dt} =  A k  \int_0^{t}
(t-t')^{k-1}
\frac{I(\zeta(t'))}{v(t')}
 dt' > 0
 \end{equation}
for $k>0$ and
\begin{equation}
-\frac{\zeta_0}{\zeta^2(t) }\frac{d \zeta}{dt} =  A
\frac{I(\zeta(t))}{v(t)}
  > 0
 \end{equation}
for $k=0$.
So, we see that the supersaturation is a decreasing function of
time.

We exclude the case $k \ll 1$ from consideration here, since the
explicit solution  for $k=0$ has  been presented.

Consider at first the times less than some time $t_p$ at which the
supersaturation $\zeta$ falls to $\zeta_0(1-\epsilon)$. Then in
the leading term equation (\ref{1}) can be rewritten as
\begin{equation}\label{4}
1- \frac{\zeta(t)}{\zeta_0 }  =  (A / v(\zeta_0)) \int_0^{t} (t-t')^k
I(\zeta(t'))
 dt'
 \end{equation}
 For interval $[\zeta_0 (1-\epsilon), \zeta_0]$ one can evidently
 use an approximation
 \begin{equation} \label{expap}
 I(\zeta(t)) = I(\zeta_0) \exp(\epsilon^{-1} \frac{\zeta(t) -
 \zeta_0 }{\zeta_0})
 \end{equation}
 Then
\begin{equation}
1- \frac{\zeta(t)}{\zeta_0 }  =  (A I(\zeta_0)/ v(\zeta_0)) \int_0^{t} (t-t')^k
\exp(-\epsilon^{-1}(1- \frac{\zeta(t')}{\zeta_0 }))
 dt'
 \end{equation}
 or
\begin{equation}
\psi(t) = (\epsilon^{-1} A I(\zeta_0)/ v(\zeta_0))
\int_0^{t} (t-t')^k
\exp(-\psi(t'))
 dt'
 \end{equation}
for the function
$$
\psi(t) =\epsilon(1- \frac{\zeta(t)}{\zeta_0 })
$$

 Certainly the last equation can be solved since the
 renormalization $$
 t \rightarrow (\epsilon^{-1} A I(\zeta_0)/ v(\zeta_0))^{1/(k+1)} t
 $$
and
$$
t' \rightarrow (\epsilon^{-1} A I(\zeta_0)/ v(\zeta_0))^{1/(k+1)} t'
$$
brings the last equation to equation with no parameters
\begin{equation}\label{universal}
\psi(t) =
\int_0^{t} (t-t')^k
\exp(-\psi(t'))
 dt'
 \end{equation}

Then the function $\psi$ is the universal function. Condition
$\psi(t_p) = 1$ makes the time $t_p$ the universal constant.

For $t \geq t_p$ solutions of (\ref{1}) and (\ref{universal})
practically coincides in their functional form. Consider  the following equation
\begin{equation}\label{5}
\frac{\zeta_0}{\zeta(t) } -1 =  A \int_0^{min(t,t_p)} (t-t')^k
\frac{I(\zeta(t'))}{v(t')}
 dt'
 \end{equation}
 which is truncated equation (\ref{1}).
Certainly  the solution $\zeta$ of equation (\ref{1}) practically
coincides with the solution $\zeta_{tr}$ of equation (\ref{5})
for $t \geq t_p$ and is less than $\zeta_{tr}$ for $t>t_p$. So, we
get the estimate from above for $\zeta$. Since
$I(\zeta_1)>I(\zeta_2)$ for two arbitrary $\zeta_1 >\zeta_2$, we
see that we know the estimate from above also for the rate of
nucleation $I$.

One can see that for $t>t_p$ the solution of (\ref{5}) is simple,
for integer $k$ the integral transforms into polynomial. Then one
can easily show that for all $k$ except $k \ll 1$ the solution
(\ref{5}) ensures the rapid decrease of $\zeta_{tr}$ and
$I(\zeta_{tr})$. So, we see the rapid decrease of $I(\zeta(t))$
based on the real solution of (\ref{1}). Namely, it means that
\begin{itemize}
\item
The nucleation period - the period of intensive formation of
embryos of a new phase is well defined. The lower boundary of this
period exists,  the long tail of the embryos size spectrum is
absent.
\item
The characteristic relative variation of $\zeta$ during the nucleation
period has the order $\epsilon$.
\item
The characteristic duration of the nucleation period has the order
$t_p$ and in comparison with the time of essential consumption of
the metastable phase $t_{fin}$ it has the smallness $\epsilon$.
\end{itemize}

The last property shows that the singularity
in
(\ref{3})
is really important and can not be neglected. It leads to the
very approximate results in the first approximation of the perturbation
theory.

One can see here another way to calculate the  characteristics of
nucleation. Really, due to the mentioned properties one can spread
equation (\ref{universal}) to the whole interval $[0,\infty]$ of
time, get the universal solution and calculate universal constants
$$
q_i = \int_0^{\infty} t^i \exp(\psi(t)) dt
$$
Only $q_i$ presents  all information important for evolution after
the nucleation period before the coalescense. For integer $k$ the
number of $q_i$ is finite,
for arbitrary $k$ one can fulfill decomposition
$$
(t-t')^k = t^k (1- \frac{t'}{t})^k = t^k (1 - k \frac{t'}{t} +
k(k-1)(\frac{t'}{t})^2/2 -...)
$$
The given number of terms ensures practically ideal result,
only first three $q_i$ are necessary.

This method which is very effective was originally proposed in
\cite{5}.

\section{The monodisperse essence of the first approximation}

Now we shall analyze the first approximation in order to see the
analytical structure. We can consider
equation (\ref{universal}) having rewritten it in a form
\begin{equation}\label{uni1}
\psi(t) = (k+1)
\int_0^{t} (t-t')^k
\exp(-\psi(t'))
 dt'
 \end{equation}
 which is more convenient to get some interesting
 properties.

The key equation to see the analytical structure is eq. (15) from \cite{4}.
One can see that the factor $\rho^k$ disappears from the
subintegral function and this corresponds to the monodisperse
approximation of the number of molecules in the droplets. Really,
in eq. (15) from \cite{4} only the distribution function $g$
without $\rho^k$ is integrated. The formal explanation is simple:
since the relative duration of nucleation period is small one can
consider that all droplets are formed simultaneously. But this
does not work at the nucleation period which is the most important
because here all droplets are formed.

Since the monodisperce approximation is adopted it
is necessary to  get an equation for the number of droplets $N$ in
this approximation.
In the mentioned renormalization it looks like
\begin{equation}
\frac{d N}{dx} = \tau^2 \exp(- \frac{N}{\tau^{k+1}} x^k )
\end{equation}
where
$$ \tau = (\frac{1}{k+1})^{1/(k+1)} $$
It is important that all already formed droplets (and even very
small ones) have equal contribution in the monodisperce
approximation. This leads to an error. The possible way of
correction is to use monodisperse approximation from \cite{9}.
It is constructed in the following manner: only droplets formed
until some moment of time are included in the monodisperce
approximation. It leads to equation of the following type for the rate of
nucleation which is proportional to the derivative of $N$ on $x$:
\begin{equation}\label{mono1}
I(x) = I(\zeta_0) \exp(-N_{pr}(\frac{x}{l}) x^k )
\end{equation}
$$
N_{pr} (x) = \int_0^x I_{pr} (x') dx'
$$
Here $I_{pr}$ is the precise solution of (\ref{uni1}).
Parameter $l$ means that only the part (namely $1/l$) of the whole
interval produces droplets which are taken into account. The value
of $l$ can be treated as the characteristic halfwidth of the
subintegral function $\rho^k g$, but later more simple definition
will be given.

One can substitute in the last equation the precise solution
$I_{pr}$ by the value calculated on the base of the same
monodisperce approximation. Then we come to the self-consistent
monodisperce approximation
\begin{equation}\label{mono2}
I_{sc}(x) = I(\zeta_0) \exp(-N_{sc}(\frac{x}{l}) x^k )
\end{equation}
$$
N_{sc} (x) = \int_0^x I_{sc} (x') dx'
$$

It is possible to continue the simplification of the monodisperse
approximation. One can see that  at the nucleation period the
argument of exponent in expression for $I$ increases very rapidly.
So, it is quite possible to assume that for $x$ at nucleation
period the argument $x/l$ corresponds to practically ideal
situation when the action of droplets is negligible even on the
rate of nucleation. It can be done when $l \gg 1$. Then we come to
$$
N(\frac{x}{l}) \approx N_1 (\frac{x}{l})=I(\zeta_0) \frac{x}{l}
$$

We see that the functional form of $I$ here is
\begin{equation}\label{first}
I (x) = I(\zeta_0) \exp(- \frac{x^{k+1}}{l}  )
\end{equation}
and it coincides
with the functional form in the iteration solution of
(\ref{uni1}).
The iteration procedure is defined in \cite{3} as
\begin{equation}\label{ite}
\psi_{i+1}(t) = (k+1)
\int_0^{t} (t-t')^k
\exp(-\psi_i(t'))
 dt' \ \ \ \psi_0 = 0
 \end{equation}
and leads to
$$
I_1 (x) = I(\zeta_0) \exp(- \frac{x^{k+1}}{k+1}  )
$$
It is known that this is rather good approximation for the rate of
nucleation aat least at $k = 3/2,2,3$ typical for
nucleation.

The full coincidence will be if we put $l=k+1$. This is the most
reasonable choice of $l$. One can see that here $l$ attains big values
 as it
has been assumed.

Certainly one can choose $l$ in a more sophisticated
style. The problem is to write and to solve some algebraic equation
on $l$ which can be done in more precise approximations. It is not
a real problem and here it is quite sufficient to use the simplest
$l=k+1$. In any case it can be refined.

Now it is worth to return to the problem of analytical calculation
of the mentioned approximations. Now only (\ref{first}) can be
calculated in analytical way. The way to calculate other
approximations is the following:
\begin{itemize}
\item
Consider (\ref{first}) as the base approximation. Having presented
$I$ as $I=I_1 \frac{I}{I_1} $ we decompose $\exp(\ln I - \ln I_1)$
into series of argument $(\ln I - \ln I_1)$. It is worth doing
because both $I$ and $I_1$ have exponential form. Then   we come to
expression of a type
$$
\int_0^x \exp(-x^{k+1}) * Polinomial(x) dx
$$
which can be easily reduced to the sum of $\zeta$-functions or
$\int_0^x \exp(-x^i) dx$.  The problem is solved.
\end{itemize}

The same method can be applied to calculate further iterations in
the
iteration procedure (\ref{ite}).

It seems that new approximations will work better than the old
ones.
Now we shall test all of them.

\section{Discussion}

Here we shall investigate in details all cases which were
calculated in \cite{4}, namely $k=0, 1/2,1,2$. We add here the case
$k=3$ because it corresponds to formation of embryos under the
free-molecular regime in a three-dimensional space which is the
most natural case.

The case $k=0$ has the analytical solution as the first order
differential equation. It is shown in figure 1. One can see that
here the rate of nucleation as a function of time (or of
coordinate $x$) is shown.

\includegraphics[angle=270,totalheight=8cm]{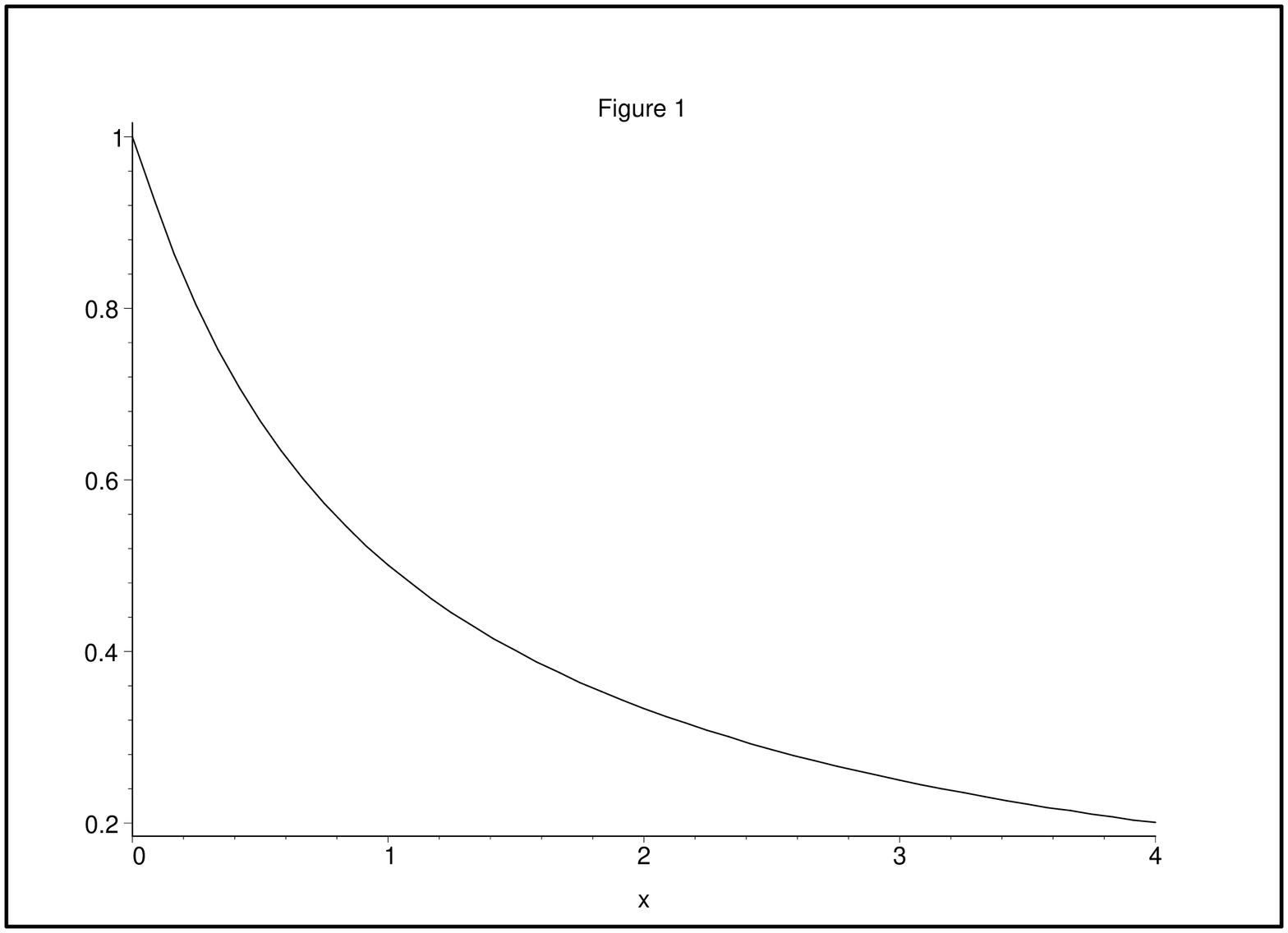}

\vspace{2cm}

The rate of nucleation as a function of $x$ forms the droplets
size spectrum. One can see that here it is rather long. The tail
plays the main role, the number of droplets is infinite at least
under the exponential approximation of the nucleation rate.

So, here appears a problem to calculate the total number of
droplets. The exponential approximation fails in this question
and this opens a problem. Fortunately, this problem has a
very simple solution.

One can see that here it is extremely simple to calculate the main
characteristic of nucleation - the total number of droplets.
Really, the total number of droplets $ N_ {tot}$
is simply the amount of
surplus substance $\zeta_0n_{\infty}$ divided by the final number of
molecules in the droplet $\nu_{fin}$:
$$
N_{tot} = \frac{\zeta_0 n_{\infty}} {\nu_{fin}}
$$
 Here we see the error of the
last model because for $k=0$ the rate of growth is zero and the
droplets can  not become supercritical ones. So, the necessity of
external parameter $\nu_{lim}$ is evident.

The case of small $k$ is the most dangerous because on one hand
there is no such balance relation as the previous one and on the
other hand all droplets play practically equal role in vapor
consumption and the number of droplets appeared at the fallen
supersaturation is essential.

Consider now the power $k=1/2$. The rates of nucleation in
different approximations in this case are drawn in figure 2.

\includegraphics[angle=270,totalheight=8cm]{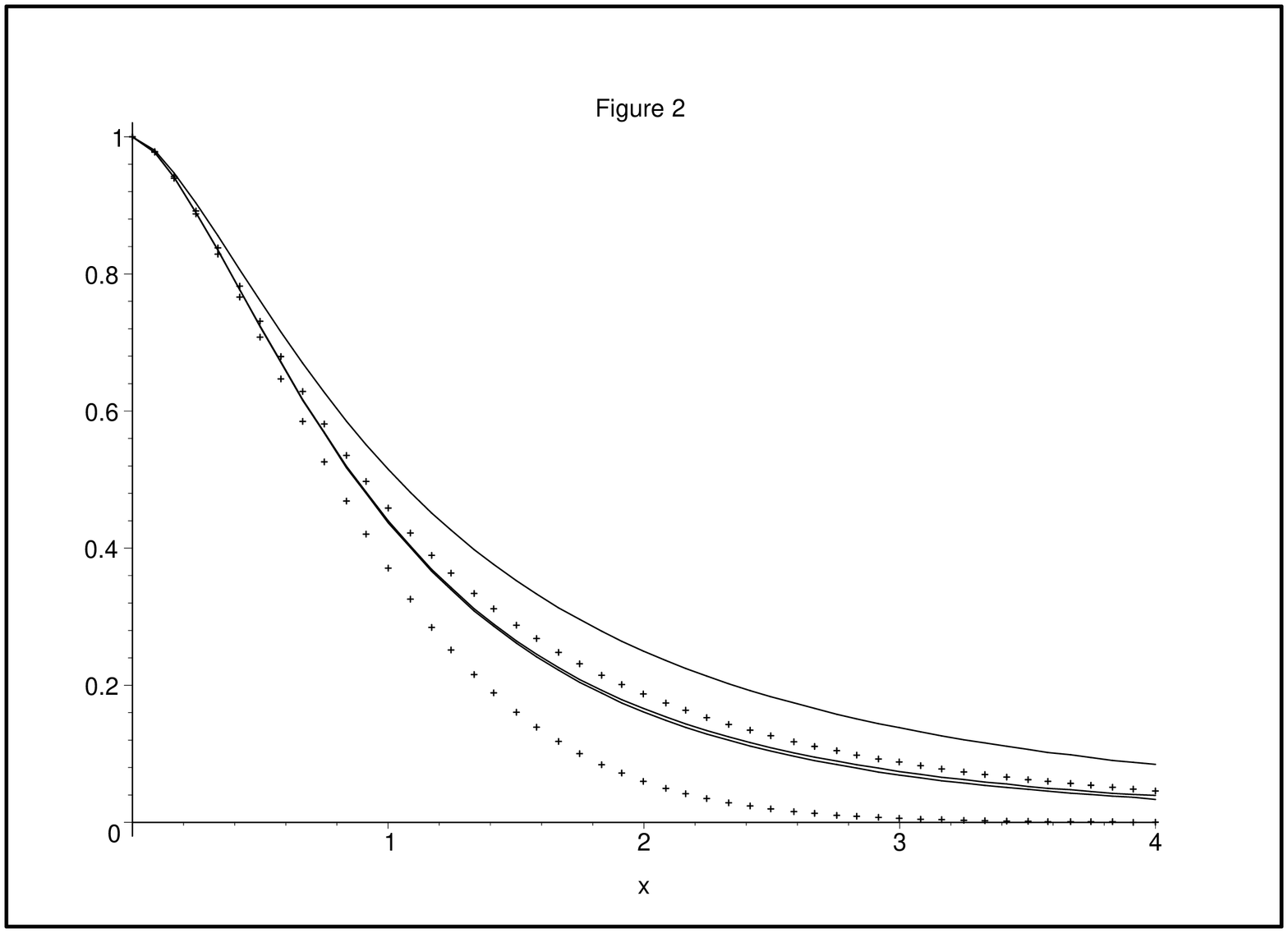}

\vspace{2cm}

The rate of nucleation is drawn.
Here one can see two point curves and two line curves. The upper
point curve is the precise solution.
Solution with decomposition
of kernel can not be separated from the real solution in scale of
a picture - the error is negligible. The upper line curve is
the approximation of the perturbation theory. It is rather close
to the real solution, but still the error exists.
 The lower
line curve is thick - two monodisperse approximations
(\ref{mono1}) and (\ref{mono2}) are drawn here. They can be hardly
separated. The approximation of the first iteration (\ref{first})
is the lower point line. One can see that here the error of the
last approximation is
essential.

Consider now the case $k=1$. This case can be solved analytically.
numerical results are drawn in figure 3.

\includegraphics[angle=270,totalheight=8cm]{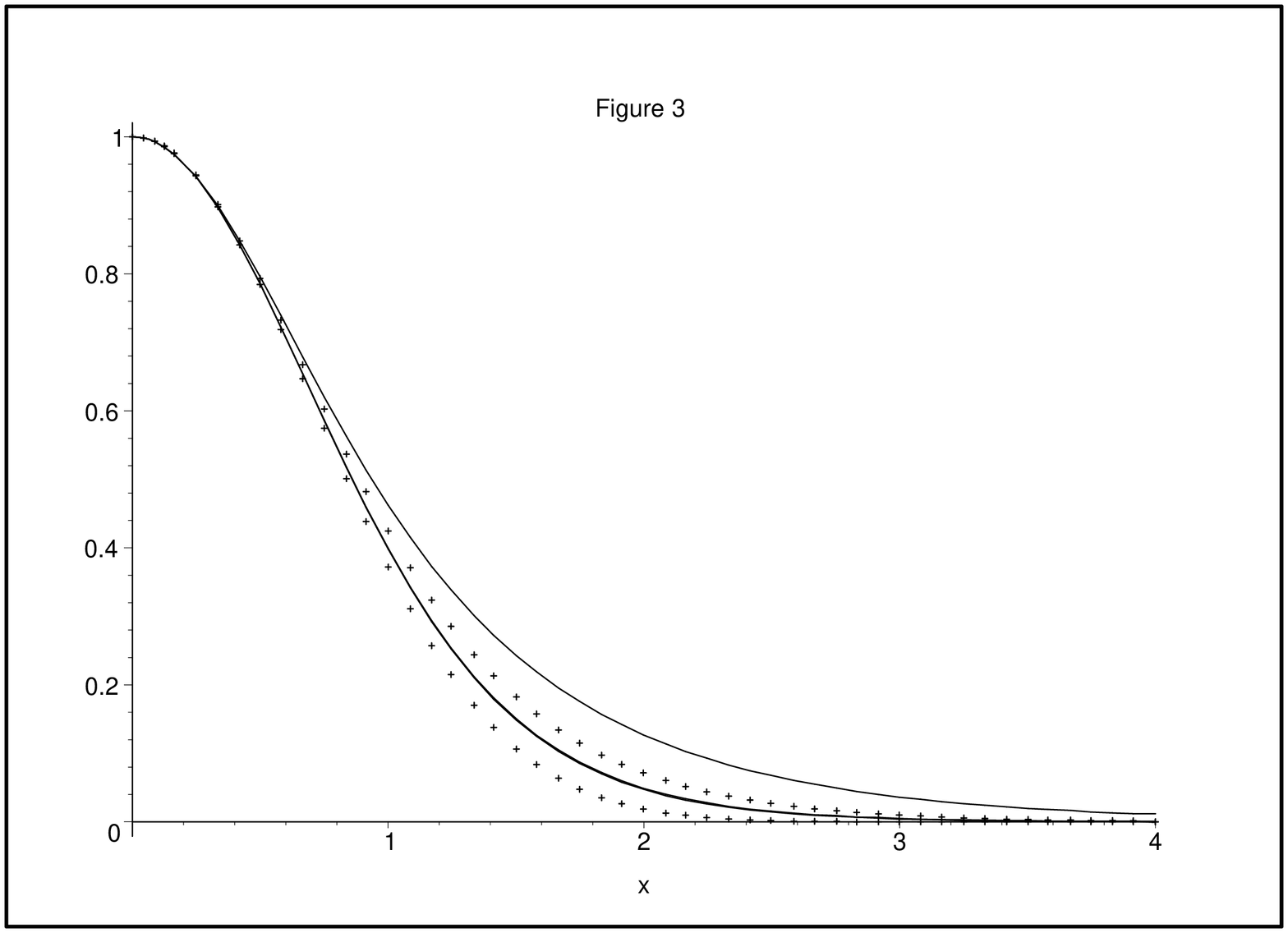}

\vspace{2cm}

The rate of nucleation is drawn.
Here one can see four curves. The upper point curve is precise
solution. The upper line curve is approximation of the
perturbation theory. The lower line curve is thick - there are two
monodisperse approximations (\ref{mono1}) and (\ref{mono2}) which
again can not be separated. The lower point curve is approximation
of the first iteration. All approximations work well but the
perturbation theory approximation has the maximum error.
Later this tendency will be stronger.

The next case is $k=2$. It is shown in figure 4.

\includegraphics[angle=270,totalheight=8cm]{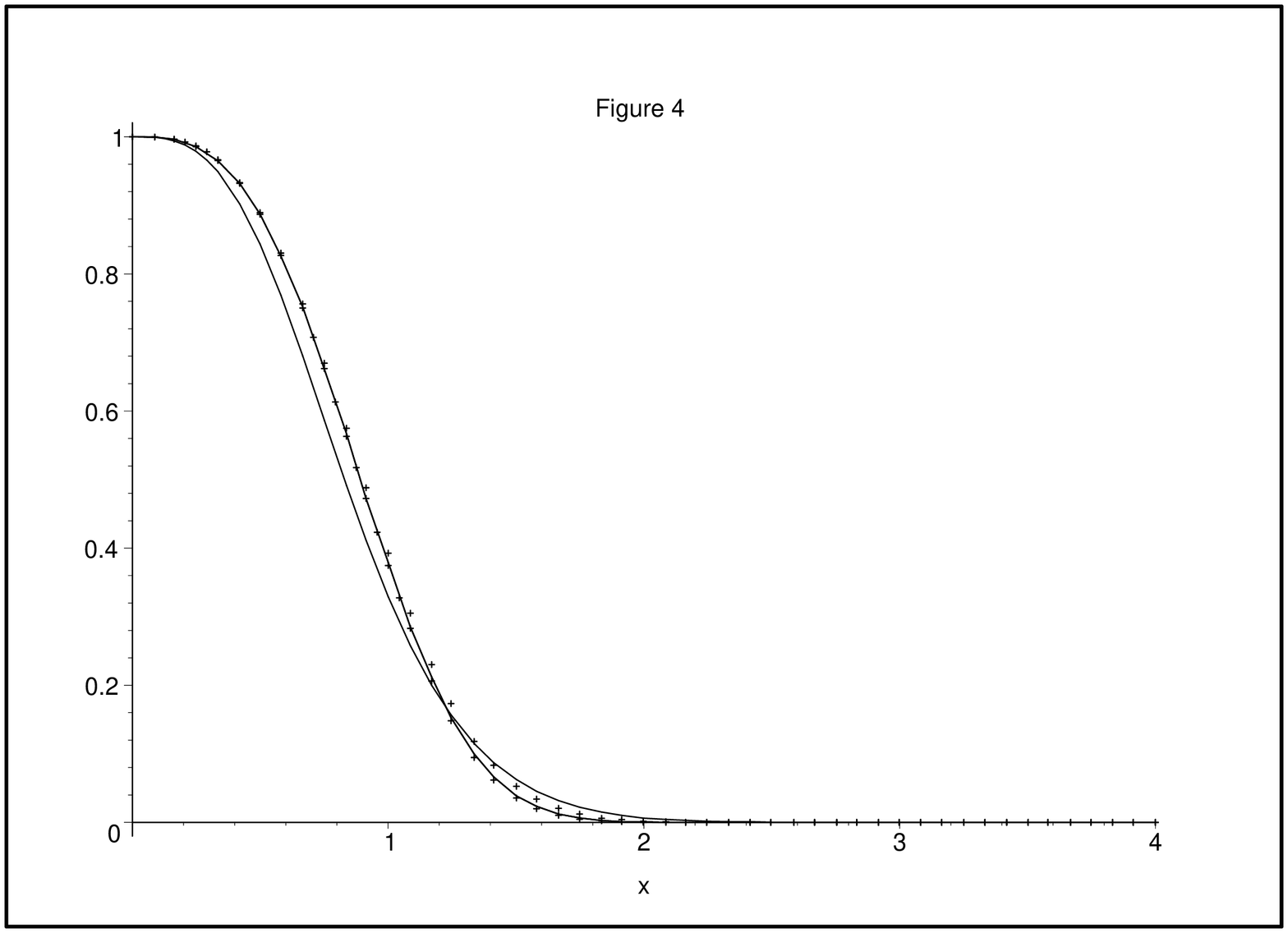}

\vspace{2cm}

The rate of nucleation is drawn.
Only two curves can be seen here.
The curves corresponding to the precise solution and
approximations (\ref{mono1}), (\ref{mono2}) and (\ref{first}) is
one thick line with points on it. Another line is the
approximation of perturbation theory. Although it has an evident
error it seems that with $k$ becoming great this error will be
small. But it is no more than illusion.

For $k=3$ we have the following numerical results which are drawn
in figure 5.

\includegraphics[angle=270,totalheight=8cm]{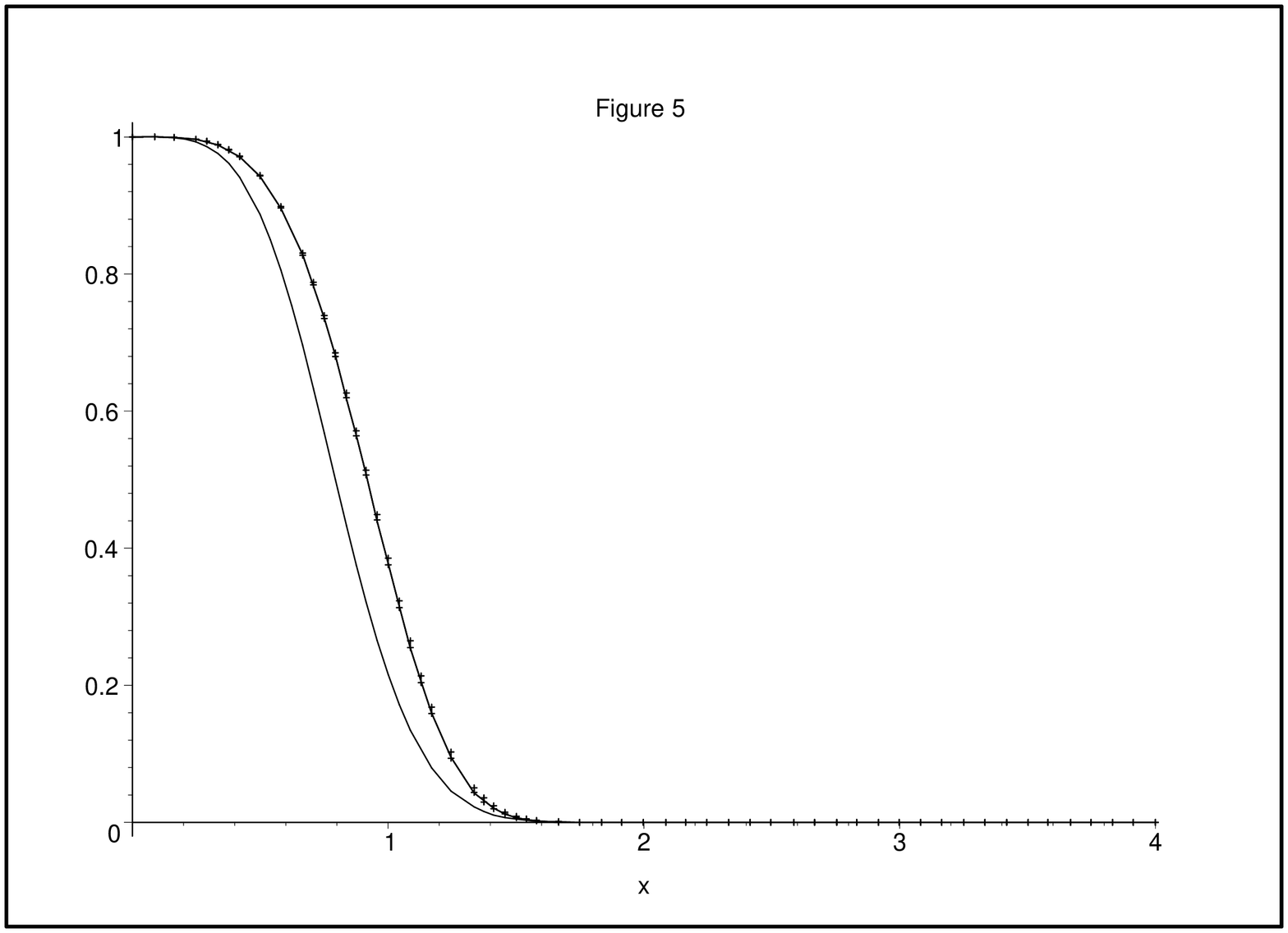}

\vspace{2cm}

The rate of nucleation is drawn.
Again one can see two curves here. The curve without points
corresponds to the approximation of the perturbation theory. The
curve with points corresponds to precise solution and all other
approximations. One can see that being compared with the case
$k=2$ the approximation of perturbation theory goes away from
precise solution. This tendency takes place because of a wrong
number of essential droplets in the actually monodisperse
approximation of the perturbation theory.

Now one can calculate the errors of new approximations. Since the
errors of the perturbation theory can be seen directly in pictures
there is no need to discuss it. The errors of approximations for
$k=1, k=2, k=3$ are drawn in figure 6.

\includegraphics[angle=270,totalheight=8cm]{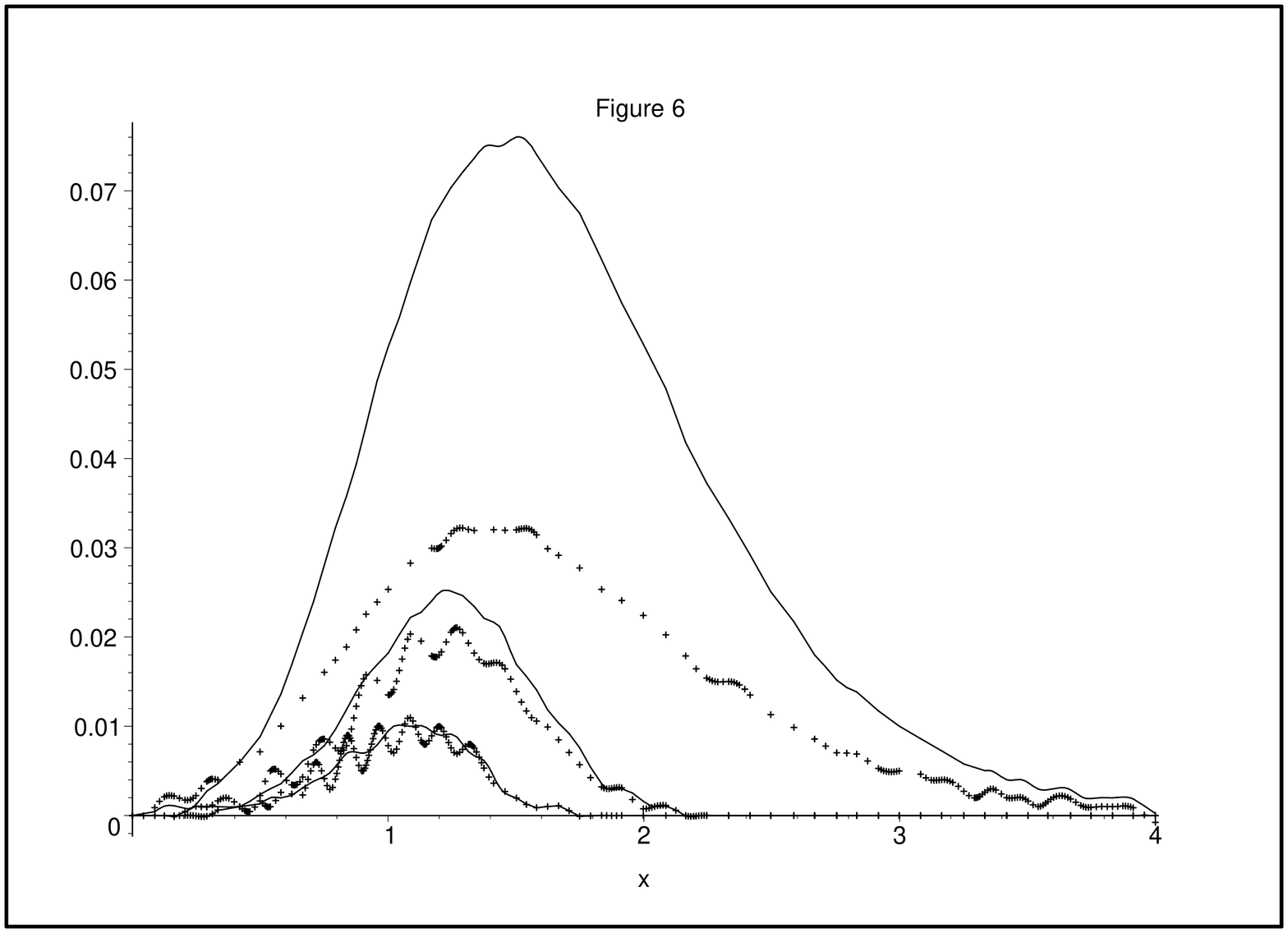}

\vspace{2cm}

One can see here three
pairs of point and line curves. The
line curves are approximations (\ref{first}) for different $k$,
the point curves are approximations (\ref{mono1}), (\ref{mono2})
(they practically coincides for all $k$ with
the reasonable accuracy of
calculations).
The lowest point and line curves
coincides (this corresponds to $k=3$) and it means that for $k>3$
there is no advantage of approximations (\ref{mono1})
instead of (\ref{first}).
The upper curves correspond to $k=1$, the middle curves - to
$k=2$. One can see that for $k=1$ the advantage of (\ref{mono1}),
(\ref{mono2}) is essential.

The approximation (\ref{first}) can lead to some  important
consequences. Having drawn in figure 7 the approximation
(\ref{first}) for
all considered $k=1/2,1,2,3$
we see that there is a focus - a point where
all curves get together.

\includegraphics[angle=270,totalheight=8cm]{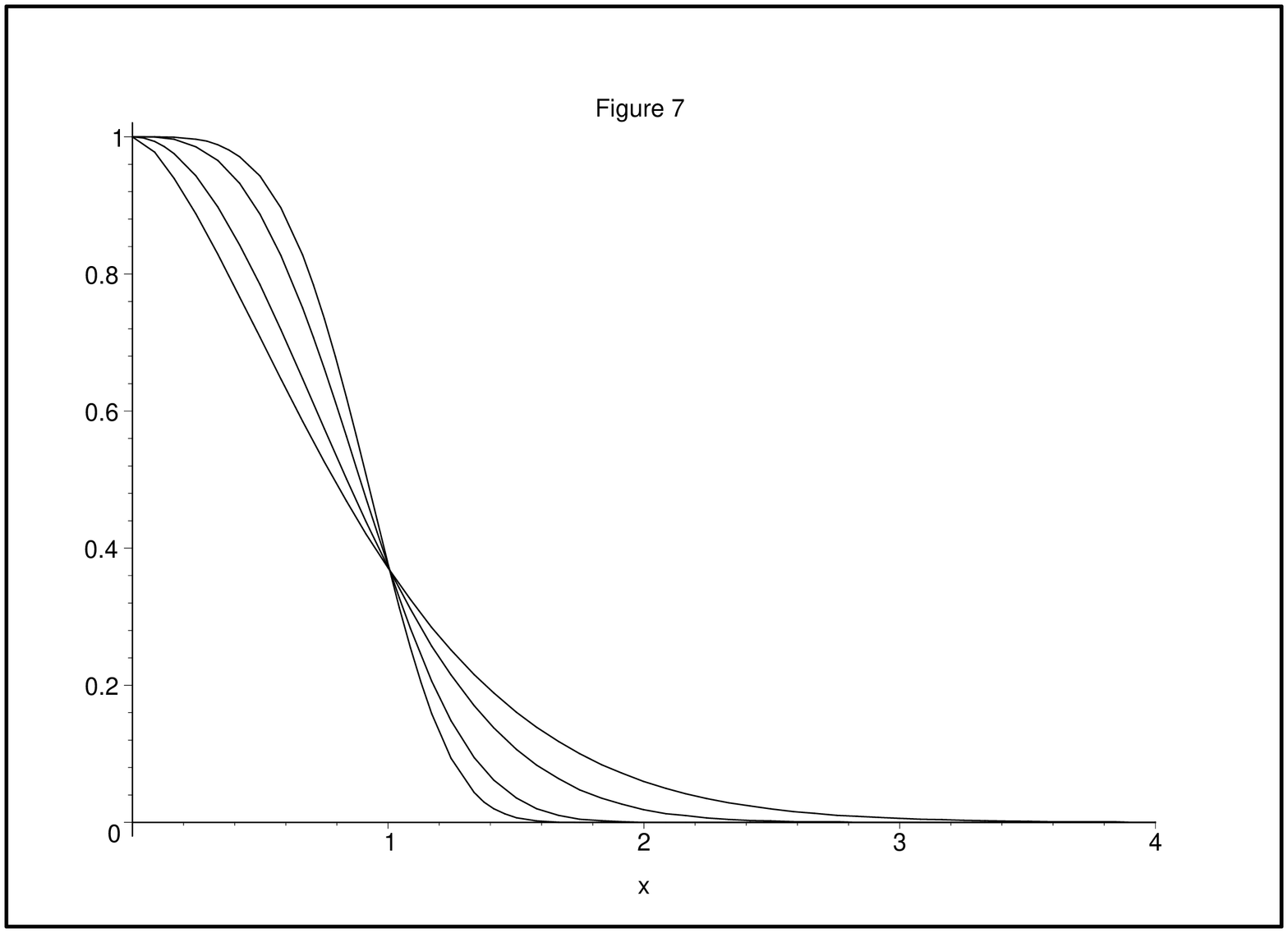}

\vspace{2cm}

This fact can be proven analytically by the simple
differentiation.  Really, the form of the first iteration is
extremely simple $\sim \exp(x^{(k+1)})$ and at $x=1$ the rate of
nucleation falls $e$-times.

Since the precise solution is not far from the first iteration,
one can see the approximate property. It is shown in figure 8
where
the precise solutions for $k=1/2,1,2,3$ are drawn.

\includegraphics[angle=270,totalheight=8cm]{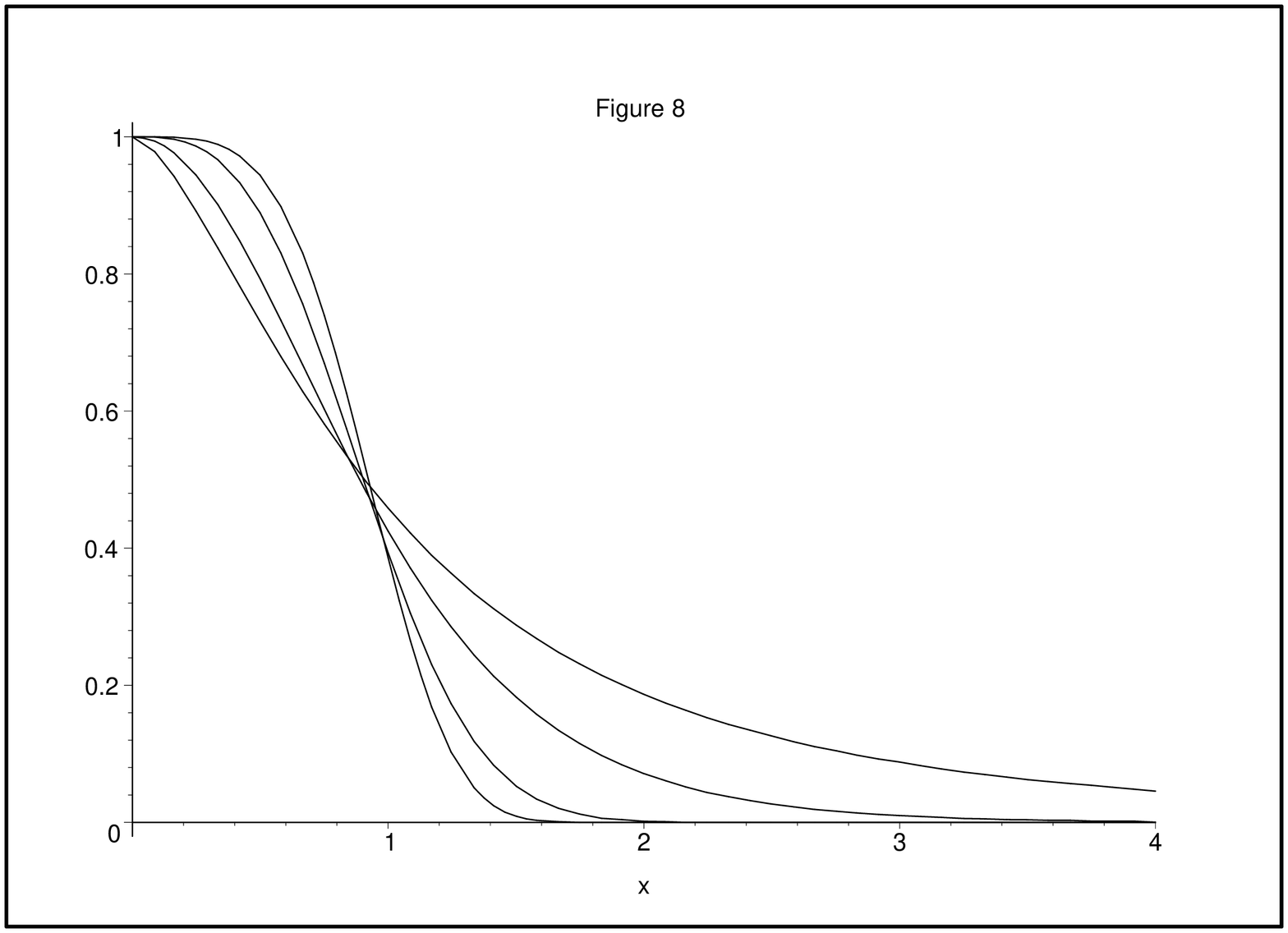}

\vspace{2cm}

This allows to speak about the property of the
precise common length of
nucleation under the external conditions of the decay type.


\begin{thebibliography}{99}
\bibitem{1}
Tunickii N.N., Journal of physical chemistry vol 15, p.
1061 (1941) (in russian)


\bibitem{2}
Kuni F.M., Grinin A.P., Kabanov A.S. Kolloidnui journal vol
46, p.440 (1984) (in russian)


\bibitem{3}
Kuni F.M., Grinin A.P. Kolloidnui journal vol
46, p.460 (1984) (in russian)


\bibitem{4}
Kukushkin S., Osipov A,
J.Chem.Phys. vol. 107 p 3247-3252  (1997)


\bibitem{5}
Kurasov V.B. Preprint of VINITI
8321-B / 5.12.1986, 40 p.

\bibitem{6}
Kurasov V.,
Kinetic theory for condensation in
dynamic conditions Physical Review E,
vol 49, p.3948-3956, 1994


\bibitem{7} Kurasov V.
Kinetic effects
of multi-component nucleation
Physica A, 8612, 2005
  58 с.

\bibitem{8}
Kuni F.M. Novojilova T.Yu. Terent'ev
I.A., Lett Math Phys. 14 161 (1987)

\bibitem{9}
Kurasov  V. Decay of metastable multicomponent mixture.
arXiv.org get
cond-mat/9310005

\end{thebibliography}
\end{document}